\documentclass[aps,prd,twocolumn,superscriptaddress,nofootinbib,longbibliography,10pt]{revtex4-2}

\usepackage{multirow}
\usepackage{amsmath}
\usepackage{amsfonts}
\usepackage{enumerate}
\usepackage{amsfonts}
\usepackage[utf8]{inputenc}
\usepackage[T1]{fontenc}

\usepackage{bbm}
\usepackage{amssymb}
\usepackage{amsthm}
\usepackage{mathtools}
\usepackage{comment}
\usepackage{mathrsfs}

\usepackage[dvipsnames]{xcolor}
\definecolor{linkcolor}{rgb}{0.0,0.3,0.5}
\usepackage[unicode, colorlinks=true, linkcolor=linkcolor, citecolor=linkcolor, filecolor=linkcolor,urlcolor=linkcolor, pdfusetitle]{hyperref}

\usepackage{hyperref}

\makeatletter
\def\@fpheader{\relax}
\makeatother


\newcounter{parentsubequation}

\makeatother

\DeclareMathAlphabet{\mathbbold}{U}{bbold}{m}{n}

\begin{document}

\title{Exceptional lines in the Kerr-Newman black hole spectrum}

\author{João Paulo Cavalcante}
\email{joao.cavalcante@ufabc.edu.br}
\author{Maurício Richartz}
\email{mauricio.richartz@ufabc.edu.br}
\affiliation{Centro de Matemática, Computação e Cognição, Universidade Federal do ABC, 09280-560, São Paulo, Brazil}

\begin{abstract}
We investigate massive scalar perturbations of Kerr-Newman black holes, focusing on the $(\ell,m) = (1,1)$ quasinormal mode spectrum in the near-extremal regime. We identify a sequence of exceptional lines, at which overtone frequencies become degenerate, together with a corresponding sequence of exceptional points for massless fields. We analyze the geometric phases associated with these exceptional points by transporting the spectrum around closed loops in parameter space and examining the resulting permutation of quasinormal mode frequencies. We further show that these degeneracies are closely related to the branching of the spectrum into zero-damping and damped modes, and derive an analytic expression for the frequencies of the zero-damping modes in the extremal limit.
\end{abstract}


\maketitle

\section{Introduction}

Black hole spectroscopy studies black holes through their characteristic oscillation frequencies~\cite{Berti:2025hly}. When a black hole is perturbed, whether by the merger of compact objects, infalling matter, or test fields, it relaxes by emitting gravitational waves~\cite{LIGOScientific:2016aoc,Dreyer_2004,PhysRevLett.27.1466,ZOUROS1979139,Vishveshwara:1970zz}. During the subsequent ringdown phase, the gravitational wave signal is dominated by a superposition of exponentially damped sinusoids, known as quasinormal modes (QNMs), which encode the fingerprint of the remnant black hole. These modes are characterized by complex frequencies whose real parts determine the oscillation frequencies and whose imaginary parts set the damping times. As the sensitivity of the LIGO-Virgo-KAGRA network continues to improve and next-generation detectors, such as LISA and the Einstein Telescope, come online, these spectral fingerprints are becoming accessible to precision measurements~\cite{LIGOScientific:2014pky, VIRGO:2014yos}.

Determining the QNM spectrum is intrinsically a non-hermitian problem. As a consequence, it can exhibit strong sensitivity to small perturbations, resonance effects, and other phenomena characteristic of non-hermitian systems~\cite{El-Ganainy:2018ksn}. In particular, exceptional points (EPs), i.e.~spectral degeneracies arising from the coalescence of distinct eigenvalues, have recently attracted considerable interest in black hole perturbation theory~\cite{Cavalcante:2024swt,PhysRevD.110.124064,mssm-ws7d,hfv8-n444,Cao:2025afs,PanossoMacedo:2025xnf,Nakamoto:2026lyo,PhysRevLett.134.141401,f8m8-vr4l}. Near EPs, the spectrum acquires a characteristic square-root branch structure, giving rise to phenomena such as mode switching, avoided crossings, and hysteresis. The latter refers to the failure of the QNM spectrum to recover its original ordering after the system parameters are varied adiabatically around a closed loop encircling one or more EPs~\cite{Cavalcante:2024swt,PhysRevD.110.124064,mssm-ws7d}.

From a gravitational perspective, degeneracies in the black hole perturbation spectrum can influence the late-time response of black holes to external perturbations. One observable consequence is that these degeneracies may leave imprints in gravitational wave ringdown signals through altered interference patterns among the excited modes \cite{Berti:2025hly, Lo:2025njp}. Near a degeneracy, the standard expansion of the ringdown signal as a sum of independent exponentially decaying QNMs acquires an additional resonant term with a linear dependence in time, marking a clear break from the conventional mode-sum picture \cite{hfv8-n444}. Recent studies further indicate that a degeneracy-based framework describes near-resonant ringdown more naturally than the traditional sum of independent damped modes, motivating a reexamination of how black hole ringdown is modeled in gravitational wave data analysis \cite{PanossoMacedo:2025xnf,pwyd-nv6v, Imafuku:2026rpn}.

The discovery of EPs in the spectrum of near-extremal Kerr black holes~\cite{Cavalcante:2024swt,PhysRevD.110.124064,mssm-ws7d} naturally raises the question of how QNM degeneracies are modified by the presence of additional physical parameters. Since EPs generally require at least a two-dimensional parameter space, higher-dimensional spaces can support richer structures, with isolated EPs organizing into lines and surfaces~\cite{ll76-j2l5}. In this context, Ref.~\cite{Cao:2025afs} recently identified a line of EPs by introducing an \textit{ad hoc} Gaussian bump in the Regge-Wheeler potential, whose amplitude, position, and width were treated as free parameters. Here, we investigate the QNM spectrum of massive scalar perturbations in Kerr-Newman (KN) black holes. The black hole charge introduces an additional parameter, enabling a search for \textit{exceptional lines} in the three-dimensional space spanned by the black hole spin, charge, and the scalar field mass. We also consider massless perturbations and investigate the existence of isolated EPs and the associated hysteresis phenomenon. Finally, we examine the connection between QNM degeneracies and the splitting of the spectrum into the zero-damping mode and damped mode branches.

This work is organized as follows. In Sec.~\ref{sec:Problem}, we review the scalar wave equation in the KN background and the associated QNM problem. In Sec.~\ref{sec:numerical}, we solve the QNM problem and scan the KN parameter space to identify exceptional lines where QNMs become degenerate. We also investigate the existence of isolated EPs and the associated hysteresis phenomenon when the scalar field is massless. In Sec.~\ref{sec:analyticFreq}, we investigate the asymptotic behavior of the QNM frequencies and their dependence on the system parameters, with particular emphasis on the extremal KN limit. Finally, Sec.~\ref{sec:remarks} summarizes our main conclusions.

Throughout this work we employ geometrical units $c=G=\hbar=1$.

\section{Linear Perturbations and Quasi-normal modes}
\label{sec:Problem}

The KN spacetime is the most general stationary and axisymmetric solution of the Einstein-Maxwell equations that describes an asymptotically flat black hole. It is characterized by three parameters: the mass $M$, the specific angular momentum $a$, and the electric charge $Q$. In Boyer-Lindquist coordinates, the metric is written as~\cite{Wald:1984}
\begin{equation}
\begin{aligned}
    ds^2 = &-\frac{\Delta}{\rho^2}(dt-a\sin^2 \theta d\phi)^2 + \frac{\rho^2}{\Delta} dr^2 + \rho^2 d\theta^2 \\& + \frac{\sin^2 \theta}{\rho^2}[(r^2+a^2)d\phi - adt]^2,
\end{aligned}
\label{kerrNewman_metric}
\end{equation}
where $ \rho^2 = r^2+a^2 \cos^2 \theta$ and $\Delta = (r-r_+)(r-r_-)$, with
\begin{equation}
    r_{\pm} = M \pm \sqrt{M^2-a^2-Q^2}.
\end{equation}
For $M^2 > a^2 + Q^2$, the KN spacetime describes a non-extremal black hole, with the event horizon and the Cauchy horizon located at $r_+$ and $r_-$, respectively. 
The angular velocities $\Omega_{\pm}$ and temperatures $T_{\pm}$ associated with the horizons $r_{\pm}$ are given by
\begin{equation}
    \Omega_{\pm} = \frac{a}{2Mr_{\pm}},\quad
    T_{\pm} = \frac{r_{\pm}-r_{\mp}}{8\pi M r_{\pm}}.
\end{equation}
In the extremal case $M^2 = a^2 + Q^2$, the two horizons coincide, yielding a single horizon at $r_+ = r_- = M$. 

Linear perturbations of a scalar field $\psi$ with mass $\mu$ in the KN background are governed by the Klein-Gordon equation,
\begin{equation}
(\nabla^\nu \nabla_\nu - \mu^2)\psi = 0.
\label{eq:KGeq}
\end{equation}
To solve this equation, we employ the separation \textit{ansatz}
\begin{equation}
\psi = R(r) S(\theta) e^{im\phi}  e^{-i\omega t},
\label{eq:ansatz}
\end{equation}
which exploits the stationarity and axisymmetry of the background by decomposing the field into modes of frequency $\omega$ and azimuthal number $m \in \mathbb{Z}$. Substituting \eqref{eq:ansatz} into \eqref{eq:KGeq} leads to two ordinary differential equations: one governing the angular part $S(\theta)$ and another governing the radial part $R(r)$. 
The radial equation is
\begin{equation}
 \! \! \frac{d}{dr}\left(\Delta \frac{dR}{dr} \right) +
\left[\frac{K^2}{\Delta}-\lambda-\mu^2r^2\right] \! R=0
   \label{eq:radeq}
\end{equation} 
where $K(r)=\omega (r^2 + a^2) - am$. 
The separation constant $\lambda$ is determined by the Sturm-Liouville problem associated with the angular equation 
\begin{equation}
  \frac{d}{du} \left[(1-u^2)\frac{dS}{du}\right]+\left(\frac{m^2}{1-u^2}
    -c^2 u^2 - A \right)S=0,
  \label{eq:angeq}
\end{equation}
with $u=\cos\theta$, $c=a\sqrt{\omega^2-\mu^2}$, and $A = \lambda + 2 a m \omega
- a^2 \omega^2$. 

The radial and angular equations must be supplemented by boundary conditions. Regularity of the scalar field at the poles ($u=\pm1$) requires the angular solutions to be spheroidal harmonics
$S_{\ell m}(\theta;c)$, with $\ell\in\mathbb{N}$ satisfying $|m|\le\ell$~\cite{1967JMP.....8.2155G,PhysRevD.73.024013}. QNMs are defined by purely ingoing waves at the outer horizon $r_+$ and purely outgoing waves at spatial infinity. Hence, the boundary conditions for $R(r)$ are
\begin{equation}
  R(r) \rightarrow
  \begin{cases}
     e^{-i(\omega - m\Omega_+)r_*}, \quad r_*\rightarrow -\infty, \\
     \frac{1}{r}e^{+i\alpha r_*}, \qquad \quad  r_*\rightarrow \infty,
   \end{cases}
\label{eq:boundCond}
\end{equation}
where $\alpha=\sqrt{\omega^2-\mu^2}$ and the tortoise coordinate $r_*$ is defined by $dr_*/dr=(r^2+a^2)/\Delta$. For fixed parameters, these boundary conditions yield a countably infinite set of QNM frequencies, labeled by the overtone number $n=0,1,2,\dots$, and typically ordered by increasing $|\operatorname{Im}(\omega_n)|$ (from least to most damped). See Ref.~\cite{Berti:2025hly} for a recent review on black hole QNMs.

\section{Numerical Results}
\label{sec:numerical}

To determine the QNM frequencies of massive scalar perturbations of a KN black hole, we solve the coupled system of Eqs.~\eqref{eq:radeq} and \eqref{eq:angeq} numerically, subject to the boundary conditions discussed in the previous section. The continued fraction method developed by Leaver~\cite{10.1098/rspa.1985.0119}, and later refined by Nollert~\cite{PhysRevD.47.5253}, is the standard approach for computing QNM frequencies, and has been applied to KN black holes in Refs.~\cite{Berti:2005eb,Konoplya:2013rxa}. Although highly efficient in general, it breaks down in the extremal limit because the event horizon becomes an irregular singular point of Eq.~\eqref{eq:radeq}. Several approaches have been proposed to overcome this difficulty~\cite{Onozawa:1995vu,Richartz:2015saa}. 

In this work, we combine Leaver's method with the isomonodromic method~\cite{CarneirodaCunha:2019tia,daCunha:2021jkm,cavalcante2023isomonodromy} to investigate QNMs in the near-extremal regime. For fixed black hole and scalar-field parameters, each method computes a QNM frequency $\omega$ together with the corresponding angular eigenvalue $\lambda$. Since both methods rely on an initial guess, different choices converge to different overtones $n$. Our implementation of Leaver's method closely follows Refs.~\cite{Dolan:2007mj,Konoplya:2006br,Konoplya:2013rxa,Siqueira:2022tbc,Richartz:2024efi}. For readers unfamiliar with the isomonodromic method, we provide a brief overview of the formalism in Appendix~\ref{sec:bondcond}. Numerical codes and datasets supporting the results of this work are available in Ref.~\cite{cavalcante_2024_13961216}.

To quantify the distance from extremality, we introduce the polar variables $\mathcal{R}$ and $\Theta$ following Ref.~\cite{Dias:2021yju}:
\begin{equation}
\frac{a}{M}=\mathcal{R} \sin \Theta, \qquad \frac{Q}{M}=\mathcal{R} \cos \Theta
\label{eq:parapolar}
\end{equation}
whose inversion gives
\begin{equation}
\mathcal{R} = \sqrt{\frac{a^2}{M^2} + \frac{Q^2}{M^2}}, \qquad \tan\Theta = a/Q,
\label{eq:polarpara}
\end{equation}
where $\mathcal{R}\in[0,1]$ and $\Theta\in[0,\pi/2]$. Extremal black holes correspond to $\mathcal{R}=1$, while $\mathcal{R}=0$ gives the Schwarzschild spacetime. The angle $\Theta$ determines the relative contribution of charge and rotation, with $\Theta=0$ corresponding to the Reissner-Nordstr\"om spacetime and $\Theta=\pi/2$ to the Kerr spacetime.

\subsection{Exceptional Lines}
\label{sec:IV-A}

\begin{figure}[htb!]
    \centering
    \includegraphics[width=1.0\columnwidth]{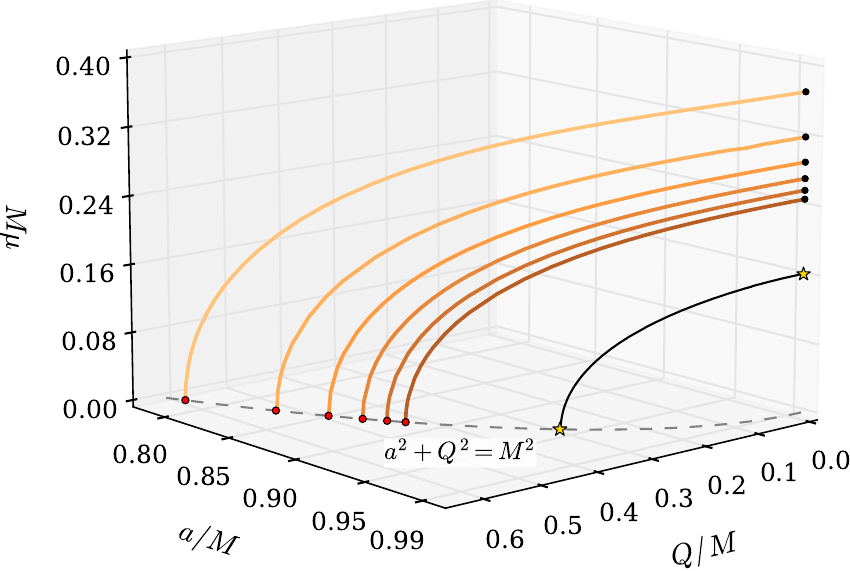}
    \caption{Exceptional lines in the parameter space $(a/M,Q/M,M\mu)$ for $(\ell,m)=(1,1)$ massive scalar perturbations of KN black holes. Each orange curve corresponds to degeneracies between the $n$-th and $(n+1)$-th QNM frequencies, with darker tones indicating higher overtones ($0\leq n\leq5$). The lines originate at the Kerr EPs in the $a/M\times M\mu$ plane (black dots) and terminate at the corresponding KN EPs in the $a/M\times Q/M$ plane (red dots). The coordinates of these endpoints are listed in Table~\ref{Tab1}. The exceptional lines lie close to, but do not intersect, the cylinder $a^2+Q^2=M^2$, which represents extremal black hole configurations. The dashed curve marks the intersection of this cylinder with the $a/M\times Q/M$ plane. The black curve, with endpoints indicated by stars, lies on the cylinder and represents the accumulation points approached by the exceptional lines as $n\rightarrow\infty$.}
    \label{fig:exceptional_line}
\end{figure}

We begin by characterizing the EP structure of the KN black hole. We focus on the $(\ell,m)=(1,1)$ sector, which is of particular interest in the near-extremal regime and provides a convenient setting for our analysis. Building on the Kerr analysis of Refs.~\cite{Cavalcante:2024swt,mssm-ws7d}, we investigate how the inclusion of electric charge extends isolated EPs into exceptional lines in the three-dimensional parameter space $(a/M,Q/M,M\mu)$.
To this end, we search for QNM degeneracies throughout the parameter space. In the Kerr limit ($Q/M=0$), Ref.~\cite{Cavalcante:2024swt} identified a degeneracy between the fundamental QNM ($n=0$) and the first overtone ($n=1$) at the critical values $(M\mu)^c \simeq 0.370498$ and $(a/M)^c \simeq 0.999465$. This point marks an EP of the spectrum. Starting from this EP, we gradually increase $Q/M$ and determine the corresponding values of $a/M$ and $M\mu$ for which the two modes remain degenerate. Each degeneracy defines a point $((a/M)^c,(Q/M)^c,(M\mu)^c)$, 
and the collection of these points traces out an exceptional line in the parameter space.
The resulting exceptional line is the uppermost curve (lightest orange tone) shown in Fig.~\ref{fig:exceptional_line}. As $(Q/M)^c$ increases from zero to $0.610458$, the critical mass $(M\mu)^c$ and spin $(a/M)^c$ both decrease, terminating at $M\mu=0$ and $(a/M)^c\simeq0.791531$ in the $a/M \times Q/M$ plane. This endpoint, corresponding to $\mathcal{R} \simeq 0.999590$ and $\Theta \simeq 52.359^\circ$, marks the appearance of an EP for massless perturbations in the KN black hole.

\begin{table}[htb!]
\centering
\begin{tabular}{ccccc}
\hline\hline
 & \quad \quad \quad Initial EPs  & \quad \quad \quad \quad Final EPs  &  \\ \hline
$n$ & $ ((a/M)^c_n,\ \ \ (M\mu)^c_n)$\ & $\ ( (a/M)^c_n,\ \ \ (Q/M)^c_n)$ & $ \mathcal{R}_n$ \\ \hline
$0$ & $\ (0.999465,\ 0.370498)$ & $\ (0.791531,\ 0.610458)$ & $\ 0.999590 $\\
$1$ & $\ (0.999853,\ 0.319135)$ & $\ (0.835717,\ 0.548945)$ & $\ 0.999881  $\\
$2$ & $\ (0.999935,\ 0.290346)$ & $\ (0.859153,\ 0.511613)$ & $\ 0.999946  $\\
$3$ & $\ (0.999963,\ 0.271509)$ & $\ (0.873924,\ 0.486000 )$ & $\ 0.999969  $\\
$4$ & $\ (0.999977,\ 0.258051)$ & $\ (0.884205,\ 0.467056 )$ & $\ 0.999979  $\\
$5$ & $\ (0.999984,\ 0.247874)$ & $\ (0.891803,\ 0.452393 )$ & $\ 0.999986  $\\
\hline
\hline
\end{tabular}
\caption{Initial and final coordinates of the endpoints of the exceptional lines shown in Fig.~\ref{fig:exceptional_line}. The initial points correspond to EPs in the Kerr limit ($Q/M=0$), while the final points correspond to EPs when the scalar field is massless ($M\mu=0$). Values of $\mathcal{R}_n$ corresponding to the final EPs are shown in the last column. For the initial EPs, one has $\mathcal{R}_n = (a/M)^c_n$.}
\label{Tab1}
\end{table}

\begin{figure*}[htb!]
  \begin{center}
\centering
\includegraphics[width=0.95\textwidth]{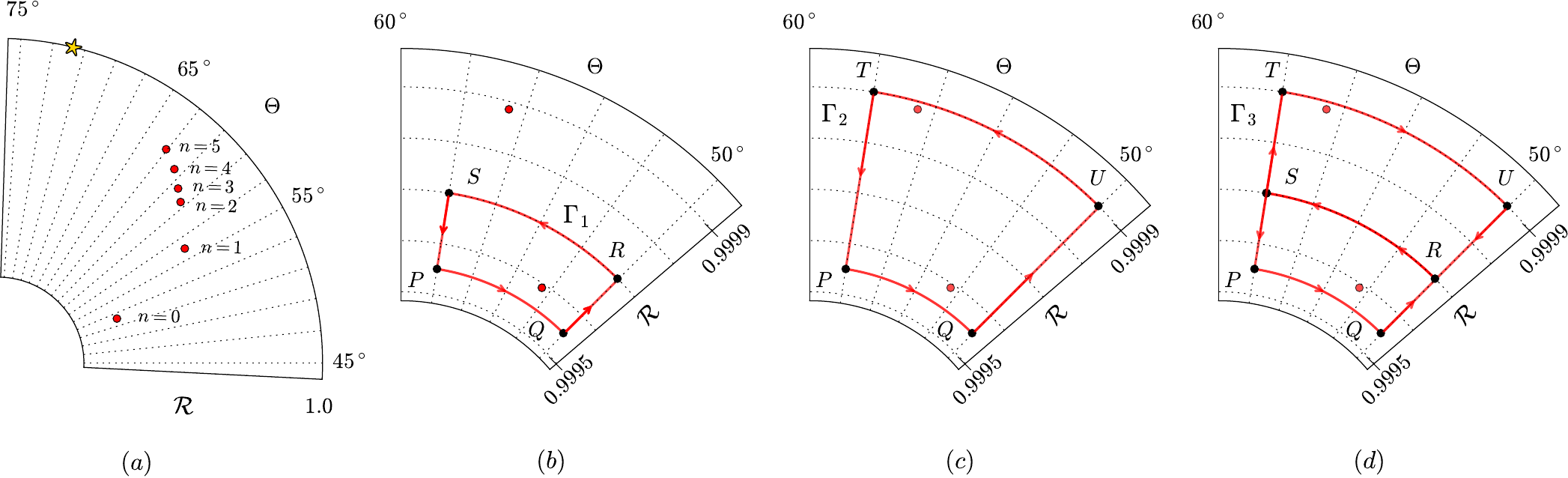}
    \caption{
    Panel (a) shows the EPs (red dots) in the $(\mathcal{R},\Theta)$ plane, corresponding to the endpoints of the exceptional lines shown in Fig.~\ref{fig:exceptional_line}. Panels (b)–(d) exhibits the paths $\Gamma_1$, $\Gamma_2$, and $\Gamma_3$. The path $\Gamma_1$, corresponding to PQRSP, encloses only the $n=0$ EP. The path $\Gamma_2$, corresponding to PQUTP, encloses the first two EPs. The path $\Gamma_3$, corresponding to PQRSTURSP, also encloses the first two EPs. All paths start at the point $\mathrm{P}=(\mathcal{R},\Theta)=(0.99955,58^\circ)$, corresponding to $(Q/M,a/M)=(0.52968,0.847666)$. For visualization purposes, the $\mathcal{R}$ axis in panel (a) is not to scale. In contrast, the $\mathcal{R}$ axes in panels (b)--(d) are shown to scale. } 
    \label{fig:ErgHys}
  \end{center}
\end{figure*}

Other EPs, labeled by the index $n$ and corresponding to degeneracies between the $n$-th and $(n+1)$-th overtones, were identified in Ref.~\cite{mssm-ws7d} for massive scalar perturbations of near-extremal Kerr black holes. Applying the same procedure described above, each of these EPs gives rise to an exceptional line in the $(a/M,Q/M,M\mu)$ parameter space. As an illustration, we consider the first six EPs listed in the first column of Table~I of Ref.~\cite{mssm-ws7d} and determine the corresponding exceptional lines by increasing the charge from the Kerr limit ($Q/M=0$).  The resulting exceptional lines are shown in Fig.~\ref{fig:exceptional_line}, where darker orange tones correspond to higher overtones. 

The initial and final points of the exceptional lines, corresponding to the EPs lying on the $a/M \times M\mu$ and $a/M \times Q/M$ planes, respectively, are indicated by the red and black dots in Fig.~\ref{fig:exceptional_line}. Their coordinates are listed in Table~\ref{Tab1}. The first column indexes the exceptional lines by the overtone index $n$, indicating a degeneracy between the $n$-th and $(n+1)$-th modes. The second column lists the EPs for massive scalar perturbations of Kerr black holes, which define the starting points (black dots) of the exceptional lines at $Q/M=0$. The third column gives the endpoints of these lines (red dots), corresponding to EPs for massless scalar perturbations of KN black holes. Finally, the last column lists the polar parameter $\mathcal{R}$, which measures the proximity of the final EPs to extremality. For the initial EPs, one simply has $\mathcal{R}_n=(a/M)^c_n$, since $(Q/M)^c_n=0$.

Note that, as the overtone number $n$ increases, the exceptional lines (and their endpoints) approach extremality. In Ref.~\cite{mssm-ws7d}, it was argued that the EPs for massive scalar perturbations of Kerr black holes form an infinite sequence accumulating at $((a/M)^c_{\star},(M\mu)^c_{\star})=(1,\ 0.16189)$, corresponding to an extremal Kerr black hole. Using the isomonodromic method, we follow this accumulation point as $Q/M$ increases from zero (see Appendix~\ref{sec8}). This yields a curve of accumulation points terminating at $((a/M)^c_{\star},(Q/M)^c_{\star})=(0.94885,\ 0.31571)$,\footnote{More precise values are $(a/M)^c_{\star}=0.9488542879718812$ and $(Q/M)^c_{\star}=0.3157143332149719$, which satisfy the extremality condition $a^2/M^2+Q^2/M^2=1$ to an accuracy of approximately $10^{-17}$. Throughout the paper, we quote only the first six significant digits.} corresponding to massless scalar perturbations of KN black holes. This curve is shown in black in Fig.~\ref{fig:exceptional_line}, with its endpoints marked by stars. It lies entirely on the cylindrical surface $a^2+Q^2=M^2$, corresponding to extremal black hole configurations in the three-dimensional parameter space.

The results presented in Fig.~\ref{fig:exceptional_line} and Table~\ref{Tab1} establish the existence of EPs for massless scalar perturbations of KN black holes and provide strong evidence for an infinite sequence of such EPs accumulating toward extremality. Figure~\ref{fig:ErgHys}(a) shows the first six EPs listed in Table~\ref{Tab1} using the polar parametrization~\eqref{eq:polarpara}. As the overtone number increases, the EPs approach the accumulation point at $\mathcal{R}=1$ and $\Theta_{\star} \simeq 71.596^\circ$.

\subsection{Hysteresis and Spectral Permutation}
\label{sec:IV-B}

We now focus on massless scalar perturbations and investigate the effect of adiabatic variations of the polar parameters $(\mathcal{R},\Theta)$. By tracing closed loops around the EPs shown in Fig.~\ref{fig:ErgHys}(a), we investigate the occurrence of hysteresis in the KN QNM spectrum, in which adiabatic transport around EPs changes the ordering of the QNM frequencies~\cite{Cavalcante:2024swt,PhysRevD.110.124064,mssm-ws7d}. We restrict our analysis to the first two EPs and consider the three loops $\Gamma_1$, $\Gamma_2$, and $\Gamma_3$, shown in Figs.~\ref{fig:ErgHys}(b)–(d). All three loops start at the point $\mathrm{P}$, corresponding to the polar coordinates $(\mathcal{R},\Theta)=(0.99955,58^\circ)$, or equivalently, to the black hole parameters $(Q/M,a/M)=(0.52968,0.847666)$.
At $\mathrm{P}$, the fundamental mode and the first two overtones, labeled by $\alpha$, $\beta$, and $\gamma$, have frequencies
\begin{subequations}\label{triplet}
\begin{align}
\omega_{\alpha} &= 0.49919 - 0.01933i, \\
\omega_{\beta}  &= 0.49643 - 0.02826i, \\
\omega_{\gamma} &= 0.49342 - 0.04514i.
\end{align}
\end{subequations}
We now determine how this triplet is transformed after one complete circuit along each loop.

Along the loop $\Gamma_1$, corresponding to the path PQRSP shown in Fig.~\ref{fig:ErgHys}(b), which encloses only the $n=0$ EP, the first two frequencies, $\omega_{\alpha}$ and $\omega_{\beta}$, are exchanged, while $\omega_{\gamma}$ remains unchanged: $(\omega_{\alpha},\omega_{\beta},\omega_{\gamma})
\rightarrow
(\omega_{\beta},\omega_{\alpha},\omega_{\gamma})$.
Reversing the orientation of $\Gamma_1$ yields the same permutation.
Along the loop $\Gamma_2$, corresponding to the path PQUTP shown in Fig.~\ref{fig:ErgHys}(c), which encloses the $n=0$ and $n=1$ EPs, the triplet undergoes the cyclic permutation
$
(\omega_{\alpha},\omega_{\beta},\omega_{\gamma})
\rightarrow
(\omega_{\gamma},\omega_{\alpha},\omega_{\beta})$.
Reversing the orientation instead gives
$
(\omega_{\alpha},\omega_{\beta},\omega_{\gamma})
\rightarrow
(\omega_{\beta},\omega_{\gamma},\omega_{\alpha})$.
Likewise, along the loop $\Gamma_3$, corresponding to the path PQRSTURSP shown in Fig.~\ref{fig:ErgHys}(d), which also encloses the first two EPs, the triplet undergoes the same cyclic permutation as for $\Gamma_2$. This shows that distinct loops can induce the same permutation. More generally, the resulting ordering depends only on the topology of the loop, namely whether it can be continuously deformed into another without crossing an EP. For an introduction to these topological concepts and their role in non-hermitian systems, see Refs.~\cite{armstrong1983basic,Guria:2023eqk,Zhong2018}.

\begin{figure}[htbp]
\centering
\begin{minipage}[b]{0.48\textwidth}
\centering
\includegraphics[width=\textwidth]{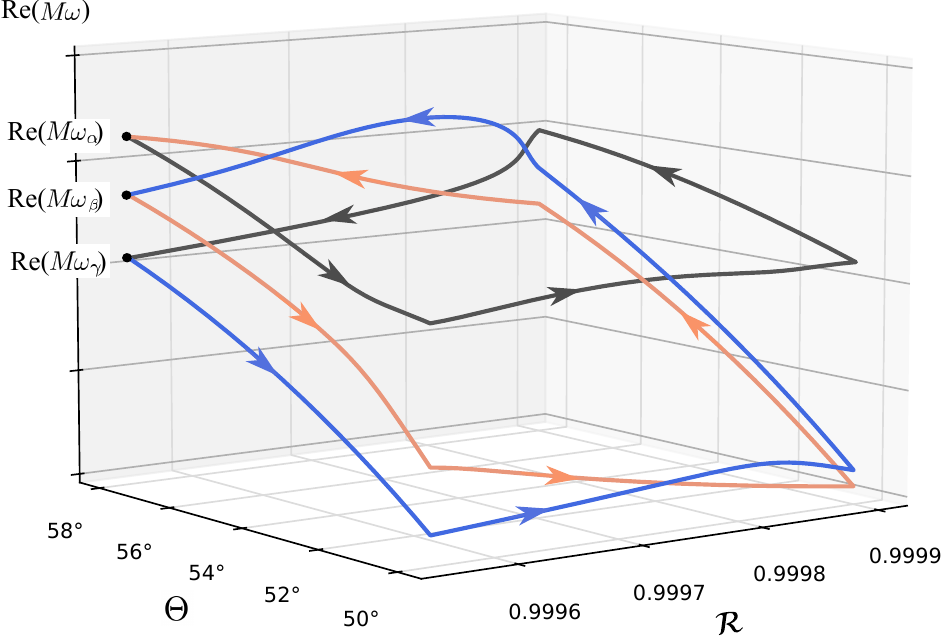}
\par\medskip
\label{fig:3Dfreq_b}
\end{minipage}
\caption{Three-dimensional visualization of the real part of the QNM frequencies, $\operatorname{Re}(M\omega)$, for the fundamental mode and the first two overtones along the path $\Gamma_2$ shown in Fig.~\ref{fig:ErgHys}(c). The horizontal axes parametrize the position along the loop in the $(\mathcal{R},\Theta)$ plane, while the vertical axis gives $\operatorname{Re}(M\omega)$. The black dots indicate the point $\mathrm{P}$ in Fig.~\ref{fig:ErgHys}(c), where the triplet of QNM frequencies $(\omega_\alpha,\omega_\beta,\omega_\gamma)$ is given by Eq.~\eqref{triplet}. The transport of the triplet around the $\Gamma_2$ loop induces the cyclic permutation $\omega_\alpha\rightarrow\omega_\gamma$, $\omega_\beta\rightarrow\omega_\alpha$, and $\omega_\gamma\rightarrow\omega_\beta$, corresponding to the gray, orange, and blue curves, respectively.}
\label{fig:3Dfreq}
\end{figure}

The hysteresis described above is characterized by a nontrivial holonomy: after completing a closed loop around one or more EPs, the frequencies $\omega_{\alpha}$, $\omega_{\beta}$, and $\omega_{\gamma}$ return to the same set of values but with a different ordering. Whether the triplet undergoes a simple exchange (as for $\Gamma_1$) or a cyclic permutation (as for $\Gamma_2$ and $\Gamma_3$) is determined solely by the topology of the loop in the parameter space punctured by the EPs. Each EP acts as a branch point, so that encircling it changes the ordering of the frequencies without altering their values. The resulting cyclic permutation is illustrated in Fig.~\ref{fig:3Dfreq}, which displays the real part of the triplet of QNM frequencies along the loop $\Gamma_2$. The different colors identify the three branches and make the cyclic permutation explicit. The imaginary parts, not shown here, exhibit the same behavior.

We emphasize that the permutation induced by a closed trajectory depends on two factors: the choice of initial modes and the set of EPs enclosed by the loop. Since the $n$-th EP couples only the $n$-th and $(n+1)$-th overtones, only modes directly connected to the enclosed EPs participate in the permutation. Consequently, because the three loops shown in Fig.~\ref{fig:ErgHys} enclose at most the first two EPs, no permutation occurs among overtones with $n\geq3$, leaving that portion of the spectrum unchanged.

\section{Branching into zero-damping modes and damped modes}
\label{sec:analyticFreq}

We now investigate how the EPs and the exceptional lines identified above are connected to the near-extremal structure of the QNM spectrum. In particular, we examine how these QNM degeneracies are related to the splitting of the spectrum into zero-damping modes (ZDMs), whose damping rates vanish as $\mathcal{R}\rightarrow 1$, and damped modes (DMs), which retain a finite damping rate in the same limit. To characterize this transition, we introduce the confluence parameter
\begin{equation}
\Lambda \simeq -\frac{i \left(M\omega \left(\cos ^2 \Theta - 2\right)+m \sin \Theta \right)}{\delta }+\mathcal{O}(\delta),
\label{confpar}
\end{equation}
with $\delta$ defined through
\begin{equation}
\cos \delta = \mathcal{R} = \sqrt{\frac{a^2}{M^2}+\frac{Q^2}{M^2}}.
\label{eq:delta}
\end{equation}
The derivation of $\Lambda$ is presented in the Appendix \ref{sec8} (following Ref.~\cite{mssm-ws7d}), while its limiting behaviour provides a convenient characterization of the QNM spectrum near extremality, distinguishing the ZDM and DM branches.

As $\delta \to 0$ (equivalently, $\mathcal{R}\rightarrow 1$), the behavior of $\Lambda$ separates two distinct regimes for a given QNM frequency. In the ZDM regime, the frequency behaves following Ref.~\cite{Dias:2021yju}:
\begin{equation} \label{eqmwre}
M\omega = m z_0 + \mathcal{O}(\delta), 
\end{equation}
with $z_0$ defined by
\begin{equation}
z_0 = \frac{\sin\Theta}{2 - \cos^2\Theta}.
\label{eq:z0}
\end{equation} 
Consequently, $\Lambda$ approaches a finite, nonzero limit. The problem is then accurately described by the confluent Heun equation, see Appendix~\ref{sec:bondcond}. On the other hand, for DMs, $M\omega$ either converges or does not converge to $m z_0$ as $\delta \to 0$. In both situations, the frequency scales nontrivially with $\delta$, leading to a divergence of the confluence parameter: $\Lambda \to \infty$. One must instead consider the doubly confluent Heun equation, see Appendix~\ref{sec8}.

\begin{figure}[htb]
    \centering
    \includegraphics[width=1.0\columnwidth]{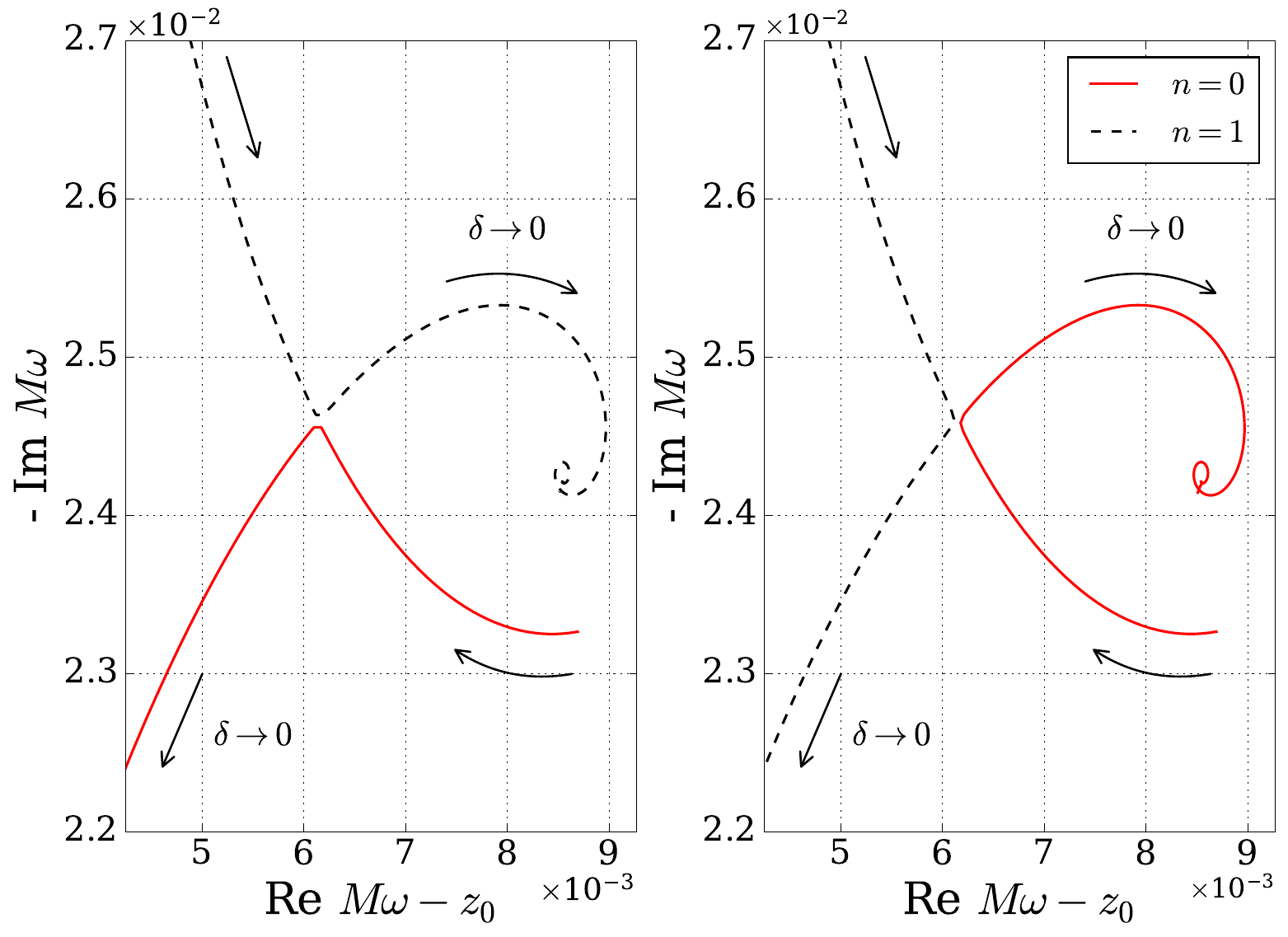}
    \caption{The pair of QNMs $(n,n+1)=(0,1)$ as a function of $a/M$, for the critical charge $(Q/M)^c_0=0.525$ and scalar field masses slightly below (left) and slightly above (right) the critical value $(M\mu)^c_0 \simeq 0.230905$. The arrows indicate increasing $a/M$ toward extremality ($\delta \to 0$). At the critical spin, $(a/M)^c_0 \simeq 0.85078$, the two modes become degenerate and subsequently exchange their character: the solid (red) branch is a ZDM and the dashed branch a DM in the left panel, whereas their roles are reversed in the right panel. The degeneracy occurs at the critical point $(a/M,Q/M,M\mu)=((a/M)^c_0,(Q/M)^c_0,(M\mu)^c_0)$, corresponding to the first EP (black dot) on the exceptional line shown in Fig.~\ref{fig:freqs}.}
    \label{fig:paraEP}
\end{figure}

Thus, the confluence parameter $\Lambda$ captures a practical way to distinguish the near-extremal behaviour of QNMs: finite $\Lambda$ signals ZDM behavior, while divergent $\Lambda$ signals DM behavior requiring a confluent analysis as $\delta \to 0$. To illustrate how the splitting of the spectrum correlates with the behavior of $\Lambda$, Fig.~\ref{fig:paraEP} shows the parametric plot of the frequencies for the fundamental mode ($n=0$) and the first overtone ($n=1$) for $(\ell,m)=(1,1)$. The two modes are shown at the critical charge, $(Q/M)_0^c = 0.525$, with the arrows indicating increasing $a/M$ toward extremality (i.e.~$\delta \to 0$). The left panel corresponds to $M\mu \lesssim (M\mu)^c_0$, while the right panel corresponds to $M\mu \gtrsim (M\mu)^c_0$, with the critical scalar field mass given by $(M\mu)^c_0 \simeq 0.230905$. At the critical spin, $(a/M)^c_0 \simeq 0.85078$, the two modes become degenerate and subsequently exchange their character. For $M\mu \lesssim (M\mu)^c_0$, the solid (red) mode is a ZDM, whereas the dashed mode is a DM. Conversely, for $M\mu \gtrsim (M\mu)^c_0$, the solid (red) mode is a DM, while the dashed mode is a ZDM. This interchange determines whether each mode is associated with a finite or divergent $\Lambda$: the divergent case gives rise to the spiral pattern shown in Fig.~\ref{fig:paraEP}, characteristic of DMs, whereas the finite case produces a nearly straight trajectory, indicating a ZDM. The transition between the DM and ZDM regimes is not unique to the $(n,n+1)=(0,1)$ pair. The same behavior is expected to occur for higher pairs of adjacent overtones, each characterized by a critical set of parameters ${(a/M)^c_n,(Q/M)^c_n,(M\mu)^c_n}$.

\begin{figure*}[htb]
  \begin{center}
\includegraphics[width=1.0\textwidth]{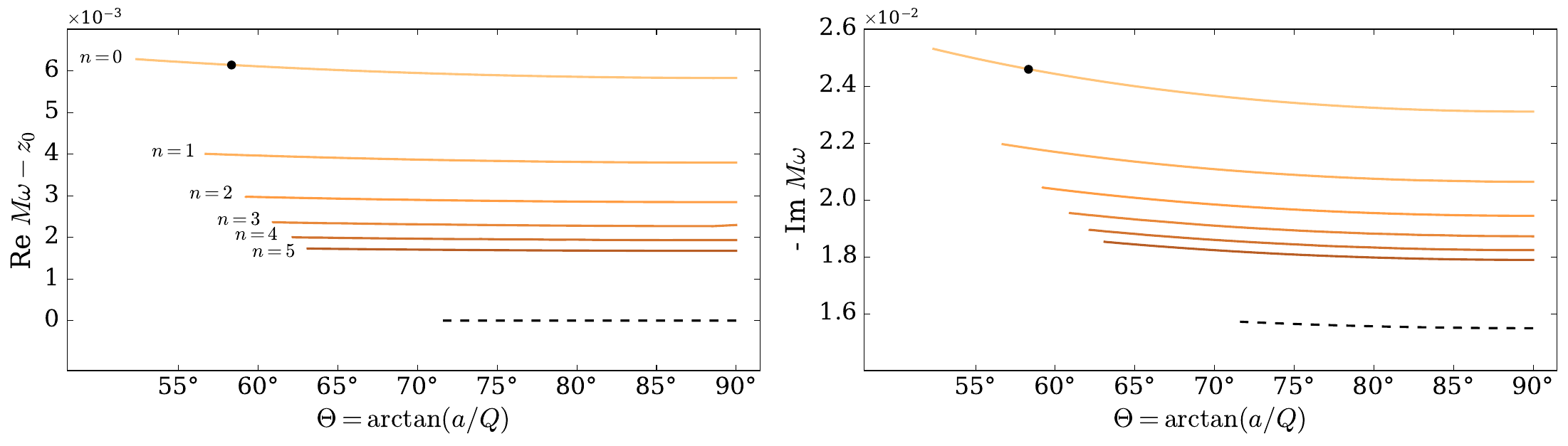}
    \caption{Behaviour of the QNM frequency along the exceptional lines of Fig.~\ref{fig:exceptional_line} as a function of $\Theta$ for the $(\ell,m)=(1,1)$ sector. As additional exceptional lines associated with higher overtones are included, the frequency of the degenerate modes approaches the extremal limit, $\mathcal{R}=1$, for $\Theta\in[\Theta_{\star},\pi/2]$ (black dashed line, $\Theta_{\star}\simeq71.596^\circ$). Along this line, $\operatorname{Re}(M\omega)=z_0=\sin\Theta/(2-\cos^2\Theta)$, while $\operatorname{Im}(M\omega)$ remains finite. The black point denotes the EP shown in Fig.~\ref{fig:paraEP}, where $M\omega-z_0\simeq0.0061-0.0245i$.} 
  \label{fig:freqs}
 \end{center}
\end{figure*}

We also investigate how the degenerate QNM frequency varies along the exceptional lines. Fig.~\ref{fig:freqs} shows the real (left) and imaginary (right) parts of the degenerate-mode frequency as functions of $\Theta$ along each exceptional line of Fig.~\ref{fig:exceptional_line}. As $n$ increases, the frequency approaches the extremal limit: $\operatorname{Re}(M\omega)-z_0$ decreases monotonically (left panel), while $\operatorname{Im}(M\omega)$ also decreases but remains finite (right panel). The black dashed curve, extending from $\Theta=\Theta_\star\simeq71.596^\circ$ to $\Theta=\pi/2$, represents the extremal limit, where $\operatorname{Re}(M\omega)=z_0$. The convergence of the real part is consistent with previous studies of ZDMs in near-extremal KN black holes, which found that $\operatorname{Re}(M\omega)$ approaches a finite limiting value at extremality~\cite{Davey:2023fin, Dias:2022oqm, Dias:2021yju}. In Fig.~\ref{fig:freqs}, however, $\operatorname{Im}(M\omega)$ remains finite along the dashed line, which corresponds to the extremal regime $\delta=0$, indicating DM behaviour and causing $\Lambda$ to diverge.

We close this section by presenting an analytic expression for the ZDM frequencies, obtained via the isomonodromic method (see Appendix~\ref{AppendAn}). For modes with finite $\Lambda$ in the extremal limit, the frequency $M\omega$ admits the following asymptotic expansion in powers of $\delta$,
\begin{equation}
\begin{aligned}
  M\omega_{k}  \simeq  \frac{m \sin\Theta}{2 - \cos^2\Theta} + 
  \left(k + \frac{1}{2} +\frac{\sigma_0}{2}\right)
  \frac{i\delta}{\cos^2\Theta-2}\\ + \left(\frac{i \sigma_1}{2\bigl(2 - \cos^2 \Theta \bigr)} - \frac{m \bigl(2 + \cos^2 \Theta \bigr) \sin \Theta}{2\bigl(2 - \cos^2\Theta\bigr)^2}\right)\delta^2, \label{eq:omegaeqoverAn}
  \end{aligned}
\end{equation}
where $k \in \mathbb{N}$, with $\sigma_0$ and $\sigma_1$ defined in Appendix~\ref{AppendAn}. The term linear in $\delta$ agrees with the expressions previously derived for massive~\cite{PhysRevD.84.044046} and massless~\cite{Hod:2008zz,Casals:2019vdb,daCunha:2021jkm} scalar fields in near-extremal black hole spacetimes (see also Refs.~\cite{1980ApJ...239..292D,Sasaki:1989ca,Glampedakis:2001js,Cardoso:2004hh,Yang:2013uba,Richartz:2017qep}). We note that the index $k$ is not, in general, equal to the overtone index $n$ of a given mode. This distinction arises because the presence of EPs (and exceptional lines) near extremality reorganizes the frequency spectrum.

\begin{figure*}[htb]
  \begin{center}
\includegraphics[width=1.00\textwidth]{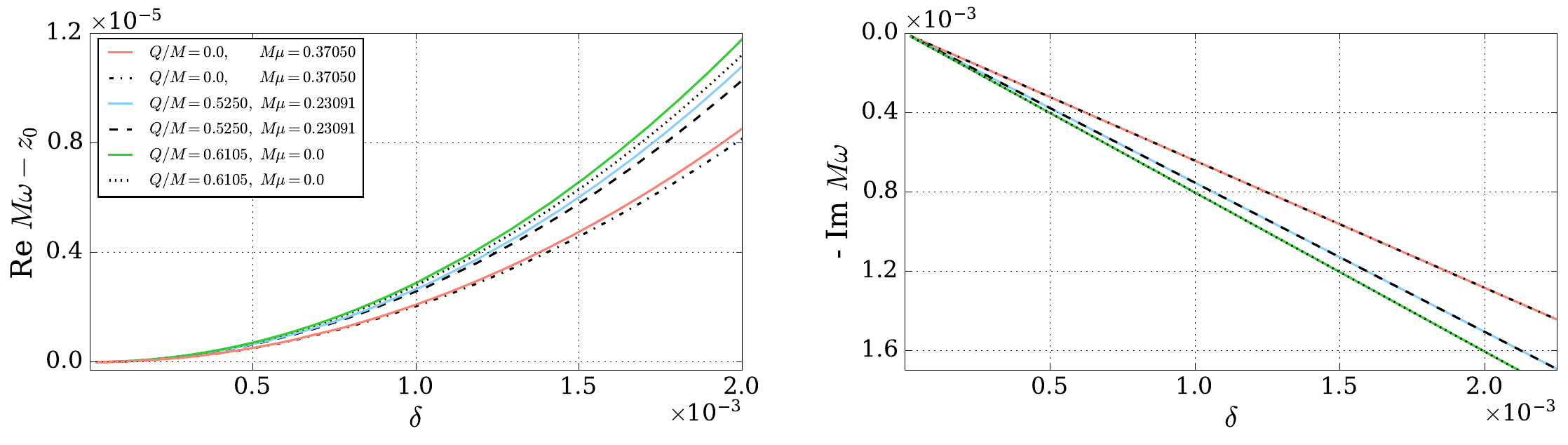}
\caption{Real (left) and imaginary (right) parts of $M\omega$ in the $(\ell,m)=(1,1)$ sector as the spin $a/M$ is varied toward extremality, plotted as functions of $\delta$, for three representative combinations of fixed $(Q/M,M\mu)$: $(0,0.37050)$, $(0.525,0.23091)$, and $(0.6105,0)$. The black curves (dashed, dot-dashed, and dotted) show the analytical near-extremal prediction of Eq.~(\ref{eq:omegaeqoverAn}), while the solid coloured curves correspond to the numerical results.}
  \label{fig:ZDMfreqs}
 \end{center}
\end{figure*}

As a validation of the analytical result in Eq.~\eqref{eq:omegaeqoverAn}, we consider three parameter sets, $(Q/M, M\mu)=(0,0.37050)$, $(0.525,0.23091)$, and $(0.6105,0)$, representative of a massive scalar perturbation in Kerr, a massive scalar perturbation in KN, and a massless scalar perturbation in KN, respectively. Fig.~\ref{fig:ZDMfreqs} shows the real (left) and imaginary (right) parts of $M\omega-z_0$ for the $(\ell,m)=(1,1)$ mode with $k=0$, as the spin $a/M$ is varied toward extremality (with the data plotted as functions of $\delta$). The non-solid curves correspond to the analytical prediction, while the solid curves are obtained numerically. In all three cases, the real part receives no linear correction in $\delta$, with deviations from its extremal value appearing only at higher orders, whereas the imaginary part follows the linear dependence on $\delta$ predicted by Eq.~\eqref{eq:omegaeqoverAn}. Although the contribution of higher-order terms becomes increasingly important as $\delta$ increases, the analytical prediction remains in excellent agreement with the numerical results throughout the near-extremal regime, confirming the validity of the asymptotic expression.

\section{Conclusion}
\label{sec:remarks}

In this work, we analyzed the near extremal QNM spectrum of scalar perturbations of KN black holes.
By exploring the three-dimensional parameter space $(a/M, Q/M, M\mu)$, we uncovered several non-hermitian features of the spectrum. In particular, we identified exceptional lines, along which two  overtones coalesce into a single degenerate mode. Building on this structure, we investigated the limit of massless scalar perturbations and identified a sequence of EPs that accumulates toward extremality as the overtone number increases. By considering adiabatic loops in the $a/M \times Q/M$ plane, we demonstrated the existence of hysteresis in the KN spectrum: after a closed loop around one or more EPs, QNM frequencies may transform into each other. This demonstration provides the first evidence of hysteresis for massless scalar perturbations of KN black holes, and the extension to the massive case is straightforward. We also investigated the behavior of the QNM frequencies along the exceptional lines and showed that the corresponding coalescence frequencies approach the extremal limit as the overtone number increases. This analysis revealed a close connection between QNM degeneracies and the splitting of the spectrum into DMs and ZDMs, for which we derived an analytic expression in the extremal limit.

Taken together, these results show that the KN QNM spectrum possesses a rich topological structure, characterized by EPs, exceptional lines, and path-dependent mode permutations. The recurrence of these phenomena across massive and massless scalar perturbations, together with the earlier Kerr results of Ref.~\cite{mssm-ws7d}, suggests that the accumulation of EPs near extremality is a robust feature of black hole spectra rather than an artifact of a particular system. More broadly, our work establishes KN black holes as a natural gravitational laboratory for exploring non-hermitian and topological phenomena in black hole physics. Future investigations may extend this analysis to nonlinear effects in the vicinity of EPs and explore possible observational signatures in gravitational-wave spectroscopy, potentially improving the characterization of astrophysical black holes~\cite{hfv8-n444,PanossoMacedo:2025xnf}.

\section{Data Availability}

Notebooks and supporting data for the findings presented in this work are available in Ref. \cite{cavalcante_2024_13961216}

\section*{Acknowledgements}
We are grateful to B.~Carneiro da Cunha for valuable discussions.
J. P. C. acknowledges the financial support from the São Paulo Research Foundation (FAPESP, Brazil), Process Number 2024/15921-9. M. R. acknowledges partial support from the Conselho Nacional de Desenvolvimento Científico e Tecnológico (CNPq, Brazil), Grant 315991/2023-2, and from the São Paulo Research Foundation (FAPESP, Brazil), Grant 2024/00923-6.

\appendix

\section{Isomonodromic Method and Boundary Conditions}
\label{sec:bondcond}

The isomonodromic method is a technique for computing QNM frequencies. It is based on the theory of isomonodromic deformations~\cite{ablowitz2006solitons,its2006isomonodromic}, according to which deformations in linear matrix systems with poles of arbitrary order that preserve monodromy correspond to completely integrable equations of mathematical physics ~\cite{Jimbo:1981aa,Jimbo:1981ab,Jimbo:1981ac}. This correspondence provides a rigorous basis for analyzing linear ordinary differential equations with both regular and irregular singularities, a class of equations that frequently appears in black hole perturbation theory. 

Monodromies, the central objects in the isomonodromic approach, were first applied in black hole physics to examine scattering data and QNMs in the infinite-damping limit~\cite{Motl:2003cd,Neitzke:2003mz,Castro:2013lba,Castro:2013kea}. In that limit, the QNM spectrum exhibits universal features that can be captured by monodromy matching techniques, which avoid direct numerical integration of the radial wave equation. These early works laid the groundwork for a more systematic exploitation of isomonodromic ideas in gravitational physics. Later, the isomonodromic method was applied to the calculation of QNM frequencies and scattering coefficients~\cite{Novaes:2014lha,CarneirodaCunha:2015hzd,Novaes:2018fry,CarneirodaCunha:2019tia,Amado:2020zsr,daCunha:2021jkm,Cavalcante:2021scq,Amado:2021erf,daCunha:2022ewy,cavalcante2023isomonodromy}. 

The method exploits the fact that both the radial \eqref{eq:radeq} and angular \eqref{eq:angeq} equations can be transformed, via appropriate coordinate changes, into the confluent Heun equation (CHE):
\begin{equation}
  \frac{d^2
   y}{dz^2}+\bigg[\frac{1-\theta_0}{z}+\frac{1-\theta_{t}}{z-t}
   \bigg]\frac{dy}{dz}-\bigg[\frac{1}{4}
     +\frac{\theta_{\star}}{2z}+\frac{t c_{t}}{z(z-t)}\bigg]y(z)=0,
  \label{heuneq}
\end{equation}
where the global \textit{monodromy} properties of the solutions of the CHE are encoded in two monodromy parameters, which we define as $\sigma$ and $\eta$ \cite{daCunha:2021jkm}. The Riemann-Hilbert (RH) map, which relates $\sigma$ and $\eta$ to the parameters of the CHE, can be expressed in terms of the Painlevé V tau-function $\tau_V$ through the following pair of equations \cite{daCunha:2021jkm},
\begin{equation}
\begin{aligned}
  \tau_V(\{\theta_k\};\sigma,\eta;t)&=0, \\
  t\frac{d}{dt}\text{log} \tau_V(\{\theta_k\}_{-};\sigma-1,\eta;t)
  -\frac{\theta_0(\theta_{t}-1)}{2}\! &= \! t c_{t}
  \label{eq:tauVcondinf}
\end{aligned}
\end{equation}
where, for convenience, we have defined $\{\theta_k\}=\{\theta_0,\theta_{t},\theta_\star\}$ and $\{\theta_k\}_-=\{\theta_0,\theta_{t}-1,\theta_\star+1\}$. The tau-function $\tau_V$ possesses the Painlevé property~\cite{Miwa:1981aa}, meaning that it is analytic in the complex plane except at the critical points $t=0$ and $t=\infty$. Hence, the map \eqref{eq:tauVcondinf} between the parameters of the CHE and the monodromy parameters $\sigma$ and $\eta$ is defined for all $t\neq 0,\infty$. By using the general expansion of $\tau_V$, as given in \cite{Gamayun:2013auu,2018JMP....59i1409L}, one can determine the eigenvalues of the CHE associated with specific boundary conditions, e.g., for QNMs. 

In the method, the QNM boundary conditions~\eqref{eq:boundCond} are encoded directly in the connection matrix $\mathsf{C}_t$. This matrix describes the linear transformation between local solutions of the radial equation near the outer horizon and those near infinity.
For the QNM condition~\eqref{eq:boundCond}, the parametrization of the connection matrix $\mathsf{C}_t$ is expressed in terms of the monodromy parameters $\sigma$ and $\eta$ as follows:
\begin{equation}
\begin{pmatrix}
\rho_{\infty}y_{\infty,+}(z) \\
\tilde{\rho}_{\infty}y_{\infty,-}(z)
\end{pmatrix}
=\mathsf{C}_{t}
\begin{pmatrix}
\rho_{t}y_{t,+}(z)\\
\tilde{\rho}_{t}y_{t,-}(z)
\end{pmatrix},
\end{equation}
with
\begin{equation}
\mathsf{C}_{t} 
= \begin{pmatrix}
e^{-\tfrac{i\pi}{2}\eta}
& 0 \\
0 &
e^{\tfrac{i\pi}{2}\eta}
\end{pmatrix}.\begin{pmatrix}
\zeta'_{t}
& e^{\tfrac{i\pi}{2}\eta}\zeta_{t} \\  e^{-\tfrac{i\pi}{2}\eta} & 1
\end{pmatrix}. \begin{pmatrix}
1
& -\zeta_{\infty,+} \\ -1 & \zeta'_{\infty,-}
\end{pmatrix}.
\label{eq:conM}
\end{equation}
The functions $y_{t,\pm}(z)$ and $y_{\infty,\pm}(z)$ are local solutions of Eq.~\eqref{heuneq} defined near $z=t$ and $z=\infty$, respectively, with the elements of the matrix $\mathsf{C}_{t}$ given by
\begin{subequations}
\begin{align}
\zeta_{\infty,+} &=  e^{\frac{i\pi}{2}\sigma}\sin\tfrac{\pi}{2}
(\theta_\star+\sigma), \\
\zeta'_{\infty,-}&= e^{-\frac{i\pi}{2}\sigma}\sin\tfrac{\pi}{2}
(\theta_\star-\sigma), \\
\zeta_{t}&=\sin\tfrac{\pi}{2}(\theta_{t}+\theta_0-\sigma)
\sin\tfrac{\pi}{2}(\theta_{t}-\theta_0-\sigma),  \\
\zeta'_{t}&=\sin\tfrac{\pi}{2}(\theta_{t}+\theta_0+\sigma)
\sin\tfrac{\pi}{2}(\theta_{t}-\theta_0+\sigma),
\end{align}
\end{subequations}
where $\rho_{t},\tilde{\rho}_{t},\rho_\infty,\tilde{\rho}_\infty$ are arbitrary normalization constants.
The requirement that the solutions of the radial equation be purely ingoing at $z=t$ and purely outgoing at $z=\infty$ implies that $\mathsf{C}_t$ is a lower triangular matrix, leading to the following relation between the monodromy parameters \cite{daCunha:2021jkm}:
\begin{eqnarray}
\begin{aligned}
  e^{\pi i(\eta+\sigma)}=
  \frac{\sin\frac{\pi}{2}(\theta_\star+\sigma)}{\sin\frac{\pi}{2}(\theta_\star-\sigma)}\prod_{\epsilon=\pm 1}
  \frac{\sin\frac{\pi}{2}(\theta_{t}+\epsilon \theta_0+\sigma)}{
    \sin\frac{\pi}{2}(\theta_{t}+\epsilon\theta_0-\sigma)}.
\end{aligned}
  \label{eq:quantizationV}
\end{eqnarray}

The same analysis applies to the angular equation, where an analogous connection matrix relates local solutions at the two regular singular points $u=\pm1$ in Eq.~\eqref{eq:angeq}. Imposing regularity at both points also forces this matrix to be lower triangular, ensuring that the solution is regular at the south and north poles:
\begin{equation}
y(z) \rightarrow
\begin{cases}
z^0\left(1+\mathcal{O}(z)\right), & z\rightarrow 1,\\
(z-t)^0\left(1+\mathcal{O}(z-t)\right), & z\rightarrow t,
\end{cases}
\label{eq:AngboundCond}
\end{equation}
This condition implies that the monodromy parameter satisfies~\cite{Cavalcante:2023rdy}
\begin{equation} \label{eq:AngboundCond2}
\sigma=\theta_0+\theta_t+2(\ell+1), \qquad \ell=0,1,2,\ldots.
\end{equation}
Notably, the condition for $\sigma$ is independent of $\eta$, demonstrating that the angular eigenvalue is determined solely by $\sigma$, $\{\theta_k\}$ and $t$.

To solve Eqs.~\eqref{eq:tauVcondinf}, we first identify the dictionary relating the radial and angular equations, Eqs.~\eqref{eq:radeq} and~\eqref{eq:angeq}, to the CHE~\eqref{heuneq}. This is achieved by suitable changes of variables that allow us to read off the monodromy parameters, modulus, and accessory parameter,
\[
(\{\theta_k\},t,c_t),\qquad
\{\theta_k\}=\{\theta_0,\theta_t,\theta_\star\},
\]
in terms of the black hole and scalar field parameters. To simplify the notation, we use the same symbols $(\theta_0,\theta_t,\theta_\star,t,c_t)$ for the radial and angular dictionary parameters. 

For the radial equation, we introduce the \textit{ansatz}
\begin{equation}
R(r)=(r-r_{-})^{-\frac{\theta_{0}}{2}}(r-r_{+})^{-\frac{\theta_{t}}{2}}y(z),
\end{equation}
with $z=2i\omega(r-r_{-})$,
so that $y(z)$ satisfies Eq.~\eqref{heuneq}. The corresponding dictionary is
\begin{subequations}\label{parameters12}
\begin{align}
\theta_{0} &=
-\frac{i}{2\pi T_{-}}\bigl(\omega-m \Omega_-\bigr), \\
\theta_{t} &=
\frac{i}{2\pi T_{+}}\bigl(\omega-m \Omega_+\bigr), \\
\theta_{\star} &=
\frac{2iM(2\omega^2-\mu^2)}{\alpha}, \\
t &= 2i\alpha(r_+-r_-), \\
tc_t &= \lambda + r_+^2\mu^2 - (3a^2 + r_-^2 + 3r_+^2)\omega^2 \nonumber\\
&\quad + i\alpha\bigl(r_- - r_+ - 2iam + 2i(a^2 + r_+^2)\omega\bigr) \nonumber\\
&\quad + i\frac{M(2\omega^2 - \mu^2)}{\alpha}
+ \frac{M^2(2\omega^2 - \mu^2)^2}{\alpha^2}.
\end{align}
\end{subequations}
Similarly, the angular equation can be brought to the canonical form by the transformation
\begin{equation}
S(u)=(1-u)^{-\frac{\theta_{0}}{2}}(1+u)^{-\frac{\theta_{t}}{2}}y(z),
\end{equation}
with $z=-2ia\omega(1-u)$, 
yielding the dictionary
\begin{subequations}\label{angparameters}
\begin{align}
\theta_{0} &= -m, \\
\theta_{t} &= m, \\
\theta_{\star} &= 0, \\
t &= -4a\alpha, \\
tc_t &= \lambda + 2(1-m)a\alpha + a^2\alpha^2.
\end{align}
\end{subequations}

To compute the QNM frequencies, we follow the procedure of \cite{PhysRevD.110.124064}. The QNMs are obtained by numerically solving the coupled algebraic system in Eq.~\eqref{eq:tauVcondinf}. For fixed black hole parameters ($a/M$, $Q/M$), field mass ($M\mu$), and quantum numbers ($\ell$, $m$), we determine the frequency $\omega$, the angular eigenvalue $\lambda$, and the auxiliary parameters $\sigma$ and $\eta$. The system yields an infinite discrete family of solutions labeled by the overtone number $n$.
Details on the isomonodromic method and its Julia implementation can be found in \cite{daCunha:2021jkm,Cavalcante:2021scq}, with corresponding codes and datasets provided in \cite{cavalcante_2024_13961216}.

\section{Riemann-Hilbert map for extremal KN black holes}
\label{sec8}

The extremal black hole geometry corresponds to a confluent limit of Eq.~\eqref{heuneq}. In terms of the parameters defined in the dictionary \eqref{parameters12}, this limit is expressed as:
\begin{subequations}\label{eq:conflimit}
\begin{align}
\Lambda &= \frac{\theta_t+\theta_0}{2} \nonumber\\
&\simeq
-\frac{i}{\delta}
\bigl[M\omega(\cos^2\Theta-2)+m\sin\Theta\bigr]
+\mathcal{O}(\delta), \\
\theta_{\circ} &= \theta_t-\theta_0, \\
u_0 &= \Lambda t,
\end{align}
\end{subequations}
where the behavior for $\Lambda$ is computed using the expressions for $\theta_0$ and $\theta_t$ in Eq.~\eqref{parameters12}. Note that $\Lambda \rightarrow \infty$ in the extremal limit $\delta \rightarrow 0$. When $\Lambda$ is infinite, one obtains an ordinary differential equation with two irregular singularities of Poincaré rank 1 at $z=0$ and $z=\infty$ \cite{NIST:DLMFc13}, namely the double-confluent Heun equation (DCHE):
\begin{equation}
\frac{d^2
  y}{du^2}+\bigg[\frac{2-\theta_{\circ}}{u}-\frac{u_0}{u^2}\bigg]
\frac{dy}{du}-\bigg[\frac{1}{4} 
+\frac{\theta_{\star}}{2u}+\frac{u_0c_{u_0}-u_0/2}{u^2}\bigg]y=0. 
\label{doubheuneq}
\end{equation}
The corresponding dictionary for the radial equation in the extremal black hole case is defined as 
\begin{equation}
(\{\theta_{\text{ext}}\},u_0,c_{u_0}),\qquad
\{\theta_{\text{ext}}\}=\{\theta_{\circ},\theta_\star\},
\end{equation}
with the parameters given by:
\begin{subequations}\label{eq:extpara}
\begin{align}
\theta_{\circ} &= -4iM\omega, \\
\theta_{\star} &= \frac{2iM(2\omega^2-\mu^2)}{\alpha}, \\
u_0 &= 4i\alpha(2M\omega - m), \\
\alpha &= \sqrt{\omega^2-\mu^2}, \\
u_0 c_{u_0} &=
M^2\mu^2
+2i\alpha\bigl(M\omega(\cos^2\Theta+1)-m\cos\Theta\bigr) \nonumber\\
 & -\frac{(\mu^2-2\omega^2)^2}{\alpha^2}
+\frac{\mu^2-2\omega^2}{\alpha}
-M^2\omega^2(\cos^2\Theta+6).
\end{align}
\end{subequations}
For the angular equation, the dictionary is identical to that of Eq.~\eqref{angparameters}, except that the extremality condition, $a^2=M^2-Q^2$, is imposed.

As seen in \cite{daCunha:2021jkm,Cavalcante:2021scq}, the Riemann-Hilbert map for the DCHE can be interpreted in terms of the third Painlevé transcendent, which is obtained by applying the confluent limit \eqref{eq:conflimit} to the function $\tau_V$. By making this, one has the result that the RH map \eqref{eq:tauVcondinf} is replaced by
\begin{equation}
\begin{aligned}
\tau_{III}(\{\theta_{\text{ext}}\};\sigma,\eta;u_0)&=0, \\
u_0\frac{d}{du_0}\text{log}
\tau_{III}(\{\theta_{\text{ext}}\}_{-};\sigma-1,\eta;u_0)-\frac{(\theta_{\circ}-1)^2}{2}&=
u_0 c_{u_0}
\end{aligned}
\label{eq:radialextremalsystemeqn}
\end{equation}
where $\{\theta_{\text{ext}}\}_{-} = \{\theta_{\circ}-1,\theta_{\star}-1\}$ are the paramaters of the map given in Eq.~\eqref{eq:extpara}. The isomonodromic $\tau_{III}$-function is a transcendental function with branch points at $u_0=0$ and $u_0=\infty$, first expanded near $u_0=0$ by Jimbo \cite{Jimbo1982MonodromyPA}. it appears in physics and mathematics problems, including random matrix theory and connection problems \cite{10.1093/imrn/rnu209,CHEN2010270}. Its expansion at $u_0=0$ is obtained through irregular conformal blocks or Fredholm determinants \cite{Gamayun:2013auu,daCunha:2021jkm}.

In terms of the boundary conditions, the confluent limit \eqref{eq:conflimit} is taken in the expression \eqref{eq:quantizationV}. In this case, the boundary conditions related to the DMs are also written in terms of the monodromy parameters as 
\begin{equation}
e^{i\pi\eta}=e^{-2\pi i\sigma}
\frac{\sin\tfrac{\pi}{2}(\theta_\star+\sigma)}{
	\sin\tfrac{\pi}{2}(\theta_\star-\sigma)}
\frac{\sin\tfrac{\pi}{2}(\theta_\circ+\sigma)}{
	\sin\tfrac{\pi}{2}(\theta_\circ-\sigma)}
\label{eq:quantizationIII}
\end{equation}
In our analysis of extremal DMs, we restrict our attention to the region Re$(M\omega) \geq mz_0$. This is because, although the expression above remains valid in the extremal case, the QNM frequencies computed in the non-extremal regime, which correspond to Re$(M\omega) < mz_0$ in the extremal limit, satisfy a condition that differs from Eq. \eqref{eq:quantizationV}, see Ref.~\cite{mssm-ws7d}. 

Finally, the damped modes whose frequencies satisfy Re$(M\omega)= m z_0$ while possessing a finite imaginary part define the extremal line shown in Fig.~\ref{fig:exceptional_line} 
for the $(\ell,m)=(1,1)$ sector. Along this curve, $M\mu$ varies in accordance with the extremality condition $M^2 = Q^2 + a^2$, while the real part of the mode frequency remains fixed at Re$(M\omega)=z_0$. This behavior for the DM frequency is also shown in Fig.~\ref{fig:freqs} as a function of $\Theta$ for $z_0 = \frac{\sin\Theta}{2 - \cos^2\Theta}.$

The numerical computation of the QNM frequencies in the extremal case follows a similar procedure used for the non-extremal case. We solve the coupled algebraic system in Eq.~\eqref{eq:radialextremalsystemeqn}, with fixed black hole parameters ($a/M$, $Q/M$), satisfying $M^2 = a^2+Q^2$, field mass ($M\mu$), and quantum numbers ($\ell$, $m$), then we determine the frequency $\omega$, the angular eigenvalue $\lambda$, and the auxiliary parameters $\sigma$ and $\eta$. We provide in Ref. \cite{cavalcante_2024_13961216} the scripts in Julia language and the dataset obtained for the DM frequencies of the $(\ell,m)=(1,1)$ sector.

\section{ZDMs: Analytic expression for $M\omega$}
\label{AppendAn}

In the following derivation, we consider in general $\ell=m \geq 1$. To derive an asymptotic expression for the frequency, we start by writing the following series expansions for the composite monodromy $\sigma$, the angular eigenvalue $\lambda$, and the frequency $M\omega$ in powers of $\delta$, 
\begin{subequations}\label{eq:exppar3}
\begin{align}
\sigma &= 1 + \sigma_0 + \sigma_1 \delta + \mathcal{O}(\delta^2), \\
\lambda &= \lambda_0 + \lambda_1 \delta + \mathcal{O}(\delta^2), \\
M\omega &= \beta_0 + \beta_1\delta + \beta_2\delta^2 + \mathcal{O}(\delta^3).
\end{align}
\end{subequations}
where $\beta_0 = mz_0 $~\cite{Dias:2021yju}. The coefficients of the $\sigma$ expansion can be computed from the second equation in \eqref{eq:tauVcondinf}, which yields the expansion for the accessory parameter $c_{t}$ provided in \cite{daCunha:2021jkm}. By making use of the $c_{t}$ expansion, one arrives at the following expressions for $\sigma_0$ and $\sigma_1$ in terms of $\lambda_0$, $\lambda_1$, and $\beta_1$: 
\begin{widetext}
\begin{subequations}\label{eq:alphaexpansion5b}
\begin{align}
\sigma_0 &=
\pm\frac{\sqrt{(1 + 4\lambda_0 + 4M^2\mu^2)(\cos^2\Theta - 2)^2
+2m^2(\cos2\Theta - 13)\sin^2\Theta}}
{\cos^2\Theta-2}, \\
\sigma_1 &=
\frac{2\lambda_1}{\sigma_0}
+\frac{16m\,\beta_1\sin\Theta}
{\sigma_0(\sigma_0^2-1)}
\Biggl[
M^2\mu^2-\lambda_0
+\frac{5(\lambda_0+M^2\mu^2)}{\cos^2\Theta-2}
-\frac{m^2(291-36\cos2\Theta+\cos4\Theta)\sin^2\Theta}
{8(\cos^2\Theta-2)^3}
\Biggr].
\end{align}
\end{subequations}
\end{widetext}

The explicit expressions for $\lambda_0$ and $\lambda_1$ are obtained as follows. First, we apply the dictionary for the angular equation given in Eq.~\eqref{angparameters}, together with the expansion for $c_t$ (see \cite{daCunha:2021jkm}) and the constraint~\eqref{eq:AngboundCond2} for $\sigma$. With these ingredients, we substitute the expansions for $c_t$ and $\lambda$ (Eq.~\eqref{eq:exppar3}) into the last expression of Eq.~\eqref{angparameters}. Then, by collecting terms order by order in $\delta$, we obtain
\begin{subequations}\label{eq:lambdaexp}
\begin{align}
\lambda_0 &=
\ell(\ell + 1)
+\frac{\bigl(1-2\ell(\ell+1)+2m^2\bigr)
(\beta_0^2-M^2\mu^2)\sin^2\Theta}
{(2\ell-1)(2\ell+3)}, \\
\lambda_1 &=
\frac{2\bigl(1-2\ell(\ell+1)+2m^2\bigr)
\beta_0\beta_1\sin^2\Theta}
{(2\ell-1)(2\ell+3)}.
\end{align}
\end{subequations}
On the other hand, to determine the parameters $\beta_1$ and $\beta_2$ we follow the same strategy as in \cite{daCunha:2021jkm} and use the dictionary \eqref{parameters12} for the radial equation.

We start by inverting the first equation in \eqref{eq:tauVcondinf} to obtain a series for $e^{i\pi\eta}$ in terms of $t$. We then use the quantization condition \eqref{eq:quantizationV} to eliminate the dependence on the monodromy parameter $\eta$. Assuming $0<\operatorname{Re}(\sigma)<1$, we find
\begin{equation}
  \Pi_{V}(\{\theta_k\};\sigma)e^{i\pi\eta}t^{\sigma-1} =
  \chi_{V}(\{\theta_k\};\sigma;t).
  \label{eq:zerotau5p}
\end{equation} 
The function $\Pi_{V}(\{\theta_k\};\sigma)$  is expressed in terms of ratios of gamma functions as
\begin{equation}
\begin{aligned}
\Pi_V(\{\theta_k\};\sigma)&=
\frac{\Gamma^2(2-\sigma)}{\Gamma^2(\sigma)}
\frac{\Gamma(\tfrac{1}{2}(\theta_\star+\sigma))}{
	\Gamma(1+\tfrac{1}{2}(\theta_\star-\sigma))}\times \\ & \prod_{\epsilon=\pm 1}
  \frac{\Gamma(\frac{1}{2}(\theta_{t}+\epsilon \theta_0+\sigma))}{
    \Gamma(1+\frac{1}{2}(\theta_{t}+\epsilon\theta_0-\sigma))}.
\end{aligned}
\label{eq:theta5}
\end{equation}
The function $\chi_{V}(\{\theta_k\} \sigma;t)$, which is analytic for small $t$ is expanded in powers of $t$ according to
\begin{multline}
\chi_V(\{\theta_k\};\sigma;t)
=1+(\sigma-1)\frac{\theta_\star
	(\theta_{t}^2-\theta_0^2)}{\sigma^2(\sigma-2)^2}t+{\cal O}(t^2).
\label{eq:chi5}
\end{multline}

Substituting the radial dictionary and the expansions from \eqref{eq:exppar3} into \eqref{eq:zerotau5p}, and including the contribution of $\chi_V(\{\theta_k\};\sigma;t)$, we find that the lowest-order terms in $\delta$ in Eq.~\eqref{eq:zerotau5p} yield
\begin{widetext}
\begin{equation}
\begin{aligned}
      e^{-\pi i\sigma_0}\frac{\Gamma(1-\sigma_0)^2}{
    \Gamma(1+\sigma_0)^2}
  \frac{\Gamma(\tfrac{1}{2}(1+\sigma_0+i\epsilon \beta_1))}{
    \Gamma(\tfrac{1}{2}(1-\sigma_0+i\epsilon \beta_1))}
  \frac{\Gamma(\tfrac{1}{2}(1+\sigma_0-4i\beta_0))}{
    \Gamma(\tfrac{1}{2}(1-\sigma_0-4i\beta_0))}
  \frac{\Gamma(\tfrac{1}{2}(1+\sigma_0-\gamma))}{
    \Gamma(\tfrac{1}{2}(1-\sigma_0-\gamma))}
    \bigg(4\delta\sqrt{4M^2\mu^2-\beta_0^2}\bigg)^{\sigma_0} =
    1+{\cal O}({\delta},\delta\log\delta)
\end{aligned}
  \label{eq:quantbeta}
\end{equation}
\end{widetext}
where
\begin{subequations}\label{eq:gamma}
\begin{align}
\epsilon &= 2(\cos^2\Theta-2), \\
\gamma &= \frac{2(m^2-2M^2\mu^2)}
{\sqrt{4M^2\mu^2-m^2}}.
\end{align}
\end{subequations}

Note that even though the expansion of $\chi_V$ is analytic in $\delta$, the term ${t}^{\sigma-1}$ in Eq.~\eqref{eq:zerotau5p} introduces non-analytic terms, such as $\delta\log\delta$, in the expression above. As we take the limit $\delta\rightarrow 0$ in Eq.~\eqref{eq:quantbeta}, the term $\delta^{\sigma_0}$ approaches zero if Re$(\sigma_0)>0$. In that case, the only way to satisfy Eq.~\eqref{eq:quantbeta} is if the argument of one of the gamma functions in the numerator approaches a non-positive integer (which we denote as $-k$, with $k \in \mathbb{Z}_+$). In fact, we can show that Eq.~\eqref{eq:quantbeta} is verified, to the lowest order in $\delta$, if
\begin{equation}
\beta_1=\frac{i}{\beta}(2k+1+\sigma_0).
\end{equation}

Having determined $\beta_1$, we now compute $\beta_2$ from the next order in $\delta$. Expanding the gamma functions on the left-hand side of \eqref{eq:zerotau5p} in powers of $\delta$ gives rise to contributions involving first derivatives of the gamma functions, cross-terms originating from the first-order corrections, as well as terms coming from $t^{\sigma-1}$. As $\delta \rightarrow 0$, we have verified that to eliminate divergences in the left-hand side, which arises from the first-derivative terms of the gamma functions, $\beta_2$ must satisfy
\begin{equation}
\beta_2 = \frac{i \sigma_1}{2\bigl(2 - \cos^2 \Theta \bigr)} - \frac{m \bigl(2 + \cos^2 \Theta \bigr) \sin \Theta}{2\bigl(2 - \cos^2\Theta\bigr)^2}.
\end{equation}
The detailed derivation of the condition for $\beta_2$ is available in \cite{cavalcante_2024_13961216}.

Substituting the expressions for $\beta_1$  and $\beta_2$ into the perturbative expression \eqref{eq:exppar3} for $M\omega$, we find that the frequencies of the ZDMs, indexed by the integer $k$, are approximated by
\begin{equation}
\begin{aligned}
  M\omega_{k} \simeq & \frac{m \sin\Theta}{2 - \cos^2\Theta} + \frac{i \delta (k + 1/2)}{(\cos^2\Theta-2)} + \frac{i \delta }{2(\cos^2\Theta-2)}\sigma_0\\ &+ \frac{i \sigma_1}{2\bigl(2 - \cos^2 \Theta \bigr)} \delta^2 - \frac{m \bigl(2 + \cos^2 \Theta \bigr) \sin \Theta}{2\bigl(2 - \cos^2\Theta\bigr)^2}\delta^2,  \label{eq:omegaeqover}
  \end{aligned}
\end{equation}
with $\sigma_0$ and $\sigma_1$ given in Eq.~\eqref{eq:alphaexpansion5b}. For the $(\ell,m)=(1,1)$ family of QNMs, we select the positive root for $\sigma_0$ in Eq.~\eqref{eq:alphaexpansion5b}.

\bibliography{KN_qnm.bib}

\end{document}